# Modeling of $C_2$ addition route to cage closure and formation of $C_{60}$

**Sabih D. Khan[1] and Shoaib Ahmad[2]**

[1]NILOP, PINSTECH, Islamabad, Pakistan.

E-mail: sabih@lahorebiz.com

[2]National Centre for Physics, QAU Campus, Shahdara Valley Road, Islamabad 44000, Pakistan
Email: sahmad.ncp@gmail.com

**Abstract**

To understand the phenomenon of fullerene growth during synthesis, an attempt is made to model a minimum energy growth mechanism using semi-empirical quantum mechanics code. $C_2$ addition by the close-caged and open-caged routes is modeled. The growth of $C_{20}$ fullerene, graphene and corannulene is also modeled. $C_{20}$ corannulene formation is identified as a major step in the growth process. Growth prior to this is followed by an open-caged route, while after $C_{20}$ corannulene formation the growth is progressed by close-caged route. A predicted mechanism for transition from open-caged $C_{24}$ to close-caged $C_{24}$ is also modeled. However, this transition needs to be further investigated. $C_{60}$ is the most stable fullerene for n < 60 and $C_{20}$ fullerene is energetically less stable as compared to $C_{20}$ corannulene. Other growth mechanisms could also occur in energetic environment commonly encountered in fullerene synthesis, but our purpose was to identify a minimal energy route, that is the most probable route.

**Keywords:** Buckyball growth, buckyball fragmentation, fullerene synthesis, semi-empirical quantum mechanics, MNDO.

PACS: 31.15.Ct, 81.05.Tp, 82.20.Wt

# 1. Introduction

Nanotechnology could be classified as design efforts to model and simulate various nano-sized machines and as implementation efforts to realize that these design as final fabricated devices and components. Molecular modeling is a tool which is required in both of the efforts; at the design stage it has the role of the paint brush and canvas, while in implementation effort it would allow nanotechnologists to simulate and model the sequence of atomic manufacturing by adding atoms or molecules one by one. This communication could be considered a part of the effort to re-evaluate methods of quantum physics and apply them to molecules.

# 2. Molecular Modeling Methods

In efforts to apply quantum physics to nano-scale objects a large number of techniques have been explored so far, which could be classified as static and dynamic molecular modeling. Static modelings involve attempts to solve time-independent Schrödinger wave equation, while dynamic modeling necessitates solving the time-dependent Schrödinger wave equation. Ab initio methods involve solutions with minimum approximations, while semi-empirical solutions use approximate parameterized values of several integrals, the parameterization is mostly obtained from experimental results or ab initio solutions. Quantum Monte Carlo (QMC) is yet another method which attempts to solve molecular systems stochastically, but whose results are so accurate that they are usually considered exact. When it comes to modeling even the simplest and smallest nano-clusters both ab initio and QMC requires extremely fast processing and very large storage systems, hence super computers are required. Semi-empirical methods on the contrary require less computational effort but provide results within certain tolerance. Semi-empirical modeling at the starting stages and ab initio or QMC at the final stages of nano-machine designing provides an optimum route. Drexler has proposed yet another route [1] that is to initially model molecules using molecular mechanics (a highly simplified molecular modeling method) at the initial stage while advanced techniques at the final stage. The dynamic methods on the other hand are much more computationally expensive and usually require tools like molecular dynamics (MD) [2], first principle molecular dynamics (FPMD) [3], etc. Dynamic methods take into account the molecular vibrations, rotations, etc. in addition to molecular instabilities due to electronic interactions. We have selected semi-empirical methods for our computational nanotechnology research based on its efficiency to model nano-clusters in less time and with considerable accuracy.

# 3. Parameterization based Semi-empirical Methods (MNDO, AM1, PM3)

The three most commonly used semi-empirical methods are MNDO [4], AM1 [5] and PM3 [6], which have certain common features. They are all self-consistent field (SCF) methods, they take into account electrostatic repulsion and exchange stabilization, and, in them, all calculated integrals are evaluated by approximate means. Further, they all use a restricted basis set of one $s$ orbital and three p orbitals ($p_x$, $p_y$, and $p_z$) per atom and ignore overlap integrals in the secular equation. Thus, instead of solving $|H - ES| = 0$ the expression $|H - E| = 0$ is solved, in which $H$ is the secular determinant, $S$ is the overlap matrix, and $E$ is the set of eigenvalues. These approximations considerably simplify quantum mechanical calculations on systems of chemical interest. As a result, larger systems can be studied [7].

All these semi-empirical methods contain sets of parameters. MNDO, AM1 and PM3 use single-atom parameters. Not all parameters are optimized for all methods; for example, in MNDO and AM1 the two electron one center integrals are normally taken from atomic spectra. In the list given in Table 1 [8], parameters are either optimized for a given method or the value of the parameter was obtained from experiment. Where neither information is given, the associated parameter is not used in that method. The three semi-empirical methods also use two experimentally determined constants per atom: the atomic mass of the most abundant isotope and the heat of atomization. MNDO, AM1 and PM3 belong to the family of NDDO (Neglect of Diatomic Differential Overlap) methods. In these methods all terms arising from the overlap of two atomic orbitals which are on different centers or atoms are set to zero [7].

## 4. Benchmarking for Fullerenes

Fullerene electronic structures differ from conventional carbon structure in hydrocarbons, on which most of the parameterization methods are based on; in the former case the carbon-carbon network forms a curved structure. The parameterization methods may not be entirely accurate for fullerenes. We have employed a brute force approach and compared the results of geometry optimization with those of the experimentally reported results. Table 2 presents the output of geometry optimization calculations carried out by the three parameterization methods i.e. MNDO, AM1 and PM3, and this when compared with the experimental results [9] indicates that the MNDO method is more accurate at least for this case of fullerene modeling, i.e. it has the lowest norm of 1.055545. Additionally, MNDO have been a favorite method for fullerene modeling by various researchers [10-12].

## 5. Justification for $C_2$ Insertion based Growth

$C_{60}$ growth can be modeled using $C_1$, $C_2$, $C_3$ or larger carbon molecules additions, but experimental observations from emission spectra of regenerative sooting discharge indicate $C_2$ to be an important constituent of carbonaceous plasma which lead to the formation of fullerenes [13]. Additionally Scheier et al. [14] also provides experimental evidence for the critical role of $C_2$ in fullerene fragmentation. We attempted to isolate the role of $C_2$ by modeling a hypothetical plasma with $C_2$ as the predominant species. A hypothetical model for the ingestion of two carbon atoms in a fullerene to fullerene conversion was first described by Endo and Kroto [15]. The parent fullerene $C_n$ should have somewhere on its surface a patch consisting of two pentagons linked to opposite edges of a central hexagon, the ingestion process is illustrated in figure 1(a) and figure 1(b). If the appropriate set of three rings is present, an isolated-pentagon isomer of $C_n$ can gain two atoms in this way, but the process yields what is presumably a high-energy form of $C_{n+2}$ with fused pentagons. A further rearrangement is needed to give a final product that has isolated pentagons, Stone-Wales transformation is one such mechanism [10, 16], as illustrated in figure 1(c) and figure 1(d).

## 6. Criteria for Stability

The SCF calculation gives density matrix (P) and Fock matrix (F) as the result. The total electronic energy $E_{elec}$ and core-core repulsion energy $E_{nuc}$ can be calculated as follows [7]:

$$E_{elec} = \frac{1}{2} \sum_\mu \sum_\nu P_{\mu\nu} \left( H_{\mu\nu} + F_{\mu\nu} \right)$$

$$E_{nuc} = \sum_A \sum_{B<A} E_N(A,B)$$

Where H is one-electron matrix, P is density matrix, F is Fock matrix and $E_{nuc}$ is core-core repulsion integral. The sum of these two energies gives the energy released when the ionized atoms and valence electrons combine to form a molecule. If we also add energy required to ionize atoms and valence electrons $E_{isol}$ and heat of atomization $E_{atom}$ then the total sum would be heat of formation at room temperature:

$$\Delta H_f^{298} = E_{elec} + E_{nuc} + \sum_A E_{isol}(A) + \sum_A E_{atom}(A)$$

Heat of formation is the energy released when isolated atoms combine to form a stable molecule. From thermodynamic argument it could be proved that this excess energy released is actually the binding energy of the molecule. So for case of $C_2$ molecule formation:

$$C + C \rightarrow C_2 + E \quad \Delta H$$

$$\Delta H = -E$$

E = Energy Released During Formation = Binding Energy = -HoF

$$-\frac{\Delta H}{n} = \text{Binding Energy per Atom}$$

For fullerenes in which all atoms are similar HoF per carbon atom is approximately equal to the negative of binding energy per atom. Binding energy per atom is the energy required to remove a single atom from the molecule. The more tightly an atom is bound in a molecule, the more stable a molecule is, so on the basis of this argument, one can treat HoF per atom as a criteria for judging the molecular stability. The lower the value of HoF per atom, the more stable the molecule.

## 7. Results of Molecular Modeling

$C_2$ addition route for open-caged fragments and close-caged fragments were modeled using Semi-empirical Quantum Mechanical method (MNDO) and the heat of formation per carbon atom is presented in figure 2, where open-caged and close-caged routes are plotted by filled circles and filled squares respectively. 3D rotateable images are also available in MDL MOL file format as multimedia 1 and multimedia 2, respectively. Heat of formation per carbon atom for the most stable fullerene isomers as given by the Atlas of Fullerenes are also presented in figure 2 and are plotted by unfilled squares. The corresponding 3D rotateable images are available in MDL MOL file format as multimedia 3. The growth route of $C_{20}$ fullerene and $C_{20}$ graphene by $C_2$ additions is plotted in figure 2 by filled diamonds and unfilled diamonds respectively and their 3D rotateable images are also available in MDL MOL file format as multimedia 4 and multimedia 5 respectively. For detailed analysis the growth of $C_{20}$ as a Corannulene, Fullerene or Graphene sheet are presented in more detail in figure 3, from n = 4 to n = 20. The geometry optimized structures by MNDO method for each cluster are illustrated next to each data point. To elaborate and clearly present results from study, the n = 20 up to n = 60 results are shown in figure 4. The geometry optimized structures again are plotted next to each data point. The predicted transformation route for $C_{24}$ fragment in $C_{24}$ fullerene is illustrated in figure 5. The geometry optimized structures are illustrated next to each data point and type of transformation is also indicated between each two consecutive data points. 3D rotateable images of this transformation are provided in MDL MOL file format as multimedia 6.

## 8. Growth of $C_{60}$ by Open Caged Fragments

The $C_{60}$ can grow either from $C_2$ by an open caged route or close caged route. The complete growth sequence is presented in figure 2 by filled circles, all energies are calculated by MNDO method as implemented by ArgusLab 1.3 [17]. The sequence can be divided into three distinct regions, i.e. from $C_2$ to $C_{20}$, from $C_{20}$ to $C_{24}$ and $C_{24}$ to $C_{60}$. Up to n= 18 the structure is two dimensional and all atoms are $sp^2$ hybridized. The transition from $sp^2$ to $sp^x$ takes place during $C_{18}$ to $C_{20}$ growth. The 1st sequence is illustrated in more detail in figure 3. It could be seen that as growth occurs from $C_2$ to larger molecules the Heat of Formation (HoF) per atom is decreased. A decrease in HoF indicates higher stability of the molecules, hence thermodynamics dictates the formation of larger carbon clusters from $C_2$ molecules. In the second region, i.e. $C_{20}$ to $C_{24}$ growth, HoF per atom increases during $C_{22}$ and $C_{24}$ formation. The thermodynamics tells that for growth to progress through this region some amount of energy is required. During fullerene synthesis in arc discharge [18], laser ablation [19], CVD [20] and even regenerative sooting discharge [21] a large distribution of energy exists and it could be possible that this excess energy required is obtained from the plasma. During the 3rd region the growth progresses similar to region 1, i.e. HoF per atom decreases with increasing size of cluster. As indicated in figure 4 and in the respective 3D rotateable images, at this stage open caged fragments grow with the addition of $C_2$ at the edges of the fragments, with $C_2$ axis in parallel to the tangential direction of the outer edge, until $C_{50}$ open-caged fragment is reached. From $C_{52}$ to $C_{60}$ the $C_2$'s are added with $C_2$ axis perpendicular to the tangential direction of the outer edge. It appears that $C_{20}$ to $C_{24}$ transition is a crucial sequence of growth, which if overcome leads to $C_{60}$ growth otherwise the growth may stop at corannulene. As evident from the HoF per atom of the molecular sequence, the HoF per atom decreases as number of carbon atoms is increased and $C_{60}$ is the one with lowest HoF per atom, which is in line with experimental observations that $C_{60}$ is the most abundant [22]. The growth of $C_{60}$ is carried from $C_2$ additions by lower energy path, i.e. at each subsequent stage the molecule becomes more stable as it grows until $C_{60}$.

## 9. Growth of $C_{60}$ by Close Caged Fullerene

For the sake of comparison, the HoF calculated for open-caged fragments is compared with that of close-caged fullerenes. Since a closed cage structure does not exist with 1 hexagon and 12 pentagons ($h = n/2 - 10$ Goldberg Polyhedra) [10], therefore $C_{22}$ does not exists, thus with $C_2$ addition we started from $C_{24}$ and went up to $C_{60}$. The only restrictions in the first case in modeling were that the structure should be close-caged and only hexagons and pentagons allowed. At this comparative stage not all structures were the most symmetric isomers, however, maximum attempts were carried out to make $C_2$ insertions in such a way as to yield more symmetric isomers. For the sake of a more thorough comparison, HoF of the most stable isomers as identified by Fowler [10] were also calculated. Geometry optimized structures of these close-caged systems are illustrated in figure 4, the corresponding HoF per atom are plotted against the open-caged fragments route. The growth sequence by this route needs to be further investigated in future with the transformation of less symmetric and less energetically favored isomers into more energetically favored isomers possibly by SW transformation. It could be observed quite clearly that HoF per atom of close-caged route above $C_{24}$ up to $C_{60}$ is lower than open-caged route. Thus above $C_{24}$ close-caged growth is more energetically favored than their corresponding open-caged fragments. But since in discharges there is a wide variation in plasma temperature and other experimental conditions, the slightly higher energy route i.e. open-caged fragments route might also occur but as soon as the fragments are extracted from the plasma and enter a low temperature region they might tend to convert into stable close-caged fullerenes by atomic rearrangement, hence only close-caged fullerenes are detected experimentally.

## 10. $C_{20}$ Growth

From the analysis so far it appears that the growth of $C_{60}$ below $C_{24}$ also needs to be explored further. So for comparison of $C_{20}$ corannulene growth sequence with other routes, an attempt was made to model the growth of two other rival $C_{20}$ molecules i.e. $C_{20}$ fullerene and $C_{20}$ graphene sheet. $C_{20}$ corannulene growth for n = 4 to n = 20 is given by molecules E2 till E10 in figure 3. In case of $C_{20}$ fullerene, since $C_{20}$ is the smallest known close-caged fullerene so only possible route that came immediately to mind was by open-caged route. Up to $C_{10}$ the growth sequence is similar to $C_{20}$ corannulene and is illustrated in figure 3, afterward the new $C_2$ molecules were inserted with $C_2$ axis perpendicular to the tangential axis of the outer most ring, this keeps on until the cage is closed. The sequence from $C_{12}$ to $C_{20}$ is illustrated in figure 3 by molecules F6, F7, F8, F9 and F10. The HoF per atom of the sequence is also presented in figure 3 by the data points with filled squares. $C_{20}$ graphene growth is illustrated in figure 3 and HoF per atom is given by data points with unfilled squares. The growth sequence is similar to $C_{20}$ corannulene and fullerene up to $C_8$, afterward the growth continues by a slightly different isomer represented by G5 and grows into G10 – the $C_{20}$ graphene.

A close analysis of $C_{20}$ growth in the figure 3 indicates that $C_{20}$ corannulene growth sequence appears to be the most stable growth sequence as it has the lowest HoF per atom. This is in line with other researchers work, i.e. detailed theoretical studies on $C_{20}$ fullerene using Quantum Monte Carlo have indicated that $C_{20}$ is at the cross-over of stability [23].

## 11. Final Picture

It appears that initially, based on our calculated values of HoF per atom and the thermodynamic considerations, growth is progressed by the open-caged fragments route up to $C_{20}$, and after $C_{24}$, the growth appears to be more favored by close caged fullerene route. During $C_{20}$ to $C_{24}$ a transition from open caged fragments to close caged fullerenes occur which further needs to be investigated. As a test case we have worked out on possible sequence for this transition and is illustrated in figure 5. It appears that for a $C_{20}$ corannulene to $C_{24}$ fullerene conversion the path of lower HoF per atom with increasing number of carbon atoms appears to be violated. While room is still open for working out the most energetically favored conversion route, but in a crude analysis it appears that a finite amount of energy has to be available in the environment for the jump from the curved $C_{20}$ to the close caged $C_{24}$. If however such energy is not available then $C_{20}$ corannulene might grow into some other structure like nanotubes or its growth may stop. Recently, one of us has presented growth selection criteria between fullerenes and nanotubes by continuum elasticity model and rotations of growing molecules [24]. Our current work hypothesizes yet another selection criterion between fullerenes and nanotubes, i.e. if sufficient environment is available for the transformation of $C_{20}$ into $C_{24}$ fullerene then $C_{60}$ growth will be more probabilistic.

## 12. Conclusion

Based on HoF per atom calculated computationally using MNDO method it appears to be that initially $C_{60}$ grow from $C_2$ by $C_2$ addition sequence up to $C_{20}$ by open caged fragments, a transition occurs from open caged fragments into close caged fullerenes during $C_{20}$ and $C_{24}$, after $C_{24}$ the growth progresses up to $C_{60}$ by close caged fullerenes with $C_2$ insertion mechanism. Other rival but energetically less favored growth routes do exist and could also be taking place, but experimental studies appears to be in favor of $C_{60}$ growth by close caged fullerenes. The transition between $C_{20}$ and $C_{24}$ needs to be further investigated, to identify the most energetically favored transition mechanism. $C_{60}$ is the most stable amongst all fullerenes and clusters as it have the lowest HoF per atom. $C_{20}$ fullerene is an unstable molecule as compared to $C_{20}$ corannulene.

**Acknowledgments**

The authors would like to thanks Ministry of Science and Technology, Pakistan for their funding in our project Carbon based Nanotechnology. We would also like to thank our team members Sohail Ahmad Janjua, Shahid Nawaz, Mashkoor Ahmad, Rahila Khalid, Abid Aleem and Bashir Ahmad for their continual support in our theoretical work. We would especially like to thank Dr. Talat S. Rahman (KSU, USA), Dr. Sikander M. Mirza (PIEAS, Pakistan) , Dr. Muhammad Mansha (PIEAS, Pakistan), Dr. Rashid Ahmad (PINSTECH, Pakistan), Ms. Tania Jabar (PINSTECH, Pakistan) and Dr. Mazhar Mehmood (PIEAS, Pakistan) for their helpful discussions related with molecular modeling, theoretical chemistry and thermodynamics. The appreciation and encouragement from Dr. Ghulam Murtaza, Dr. Kamal uddin and their colleagues at Pakistan Physical Society could also not be forgotten.

**References**

[1] Drexler K. E. 1992 *Nanosystems; Molecular machinery, manufacturing and computation* (Wiley Interscience) chapter 3

[2] Hinchliffe A. 1999 *Chemical Modeling; From Atoms to Liquids* (West Sussex: John Wiley & Sons Ltd.) 183-190

[3] Car R. & Parrinello M. 1985 *Physical Review Letters* **55** 2471

[4] Dewar M. J. S. & Thiel W. 1977 *JACS* **99** 4899-4907

[5] Dewar M. J. S., Zoebisch E. G., Healy E. F. & Stewart J. J. P. 1985 *JACS* **107** 3902-3909

[6] Stewart J. J. P. 1989 *J. Comp. Chem.* **10** 209-220
Stewart J. J. P. 1989 *J. Comp. Chem.* **10** 221-264

[7] Drakos N., Hennecke M., Moore R., Swan H., Lippmann J., Rouchal M. and Wilck M. 1998 MOPAC Manual (http://home.att.net/~mopacmanual/node7.html).

[8] Drakos N., Hennecke M., Moore R., Swan H., Lippmann J., Rouchal M. and Wilck M. 1998 MOPAC Manual (http://home.att.net/~mopacmanual/node441.html).

[9] Haddon R. C. 1993 *Science* **261** 1545-1550

[10] Fowler P. W. & Manolopoulos D. E. 1995 *An Atlas of Fullerenes* (New York: Oxford University Press Inc.) pages 73 120-125 15-17 76-77

[11]     Weltner Jr. W. & Zee R. J. V. 1989 *Chem. Rev.* **89** 1713-1747

[12]     Plater J., Rzepa H. S., Stoppa F. and Stossel S. 1994 *J. Chem. Soc. Perkin Trans* **2** 399

[13]     Ahmad S., Qayyum A., Akhtar M. N. & Riffat T. 2000 *Nuclear Instruments & Methods in Physics Research B* 1-7

[14]     Scheier P., Dunser B., Worgotter R., Muigg D., Matt S., Echt O., Foltin M., Mark T. D. 1996 *Physical Review Letters* **77** 2654-2657

[15]     O'Brian S. C., Heath J. R., Curl R. F. & Smalley R. E. 1988 *J. Chem. Phys.* **88** 220

[16]     Stone A. J. & Wales D. J. 1986 *Chem. Phys. Lett.* **128** 501

[17]     Thompson M. A., Zerner M. C. 1991 *J. Am. Chem. Soc.* **113** 8210
         Thompson M. A., Glendening E. D. & Feller D. 1994 *J. Phys. Chem.* **98** 10465-10476
         Thompson M. A. & Schenter G. K. 1995 *J. Phys. Chem.* **99** 6374-6386
         Thompson M. A. 1996 *J. Phys. Chem.* **100** 14492-14507

[18]     Kratschmer W., Lamb L. D., Fostiropoulos K. & Huffman D. R. 1990 *Nature* **347** 354

[19]     Ying Z. C., Hettich R. L., Compton R. N. and Haufler R. E. 1996 *J. Phys. B: At. Mol. Opt. Phys* **29** 4935-4942

[20]     Dresselhaus M. S., Dresselhaus G., Avouris P. (Eds.) 2001 Topics Appl. Phys. **80** 32-44

[21]     Ahmad S. 2001 *Eur. Phys. J. D* **15** 349-354

[22]     Kroto H. W., Heath J. R., O'Brien S. C., Curl R. F. & Smalley R. E. 1985 *Nature* **318** 162

[23]     Kent P. R. C., Towler M. D., Needs R. J. and Rajagopal G. 2000 *Physical Review B* **62** 15394

[24]     Ahmad S. 2005 *Nanotechnology* **16** 1739

**Figure Captions**

Figure 1. Fullerenes may grow by $C_2$ insertion which introduces a hexagon with every insertion, the unstable isomers may then be converted into stable ones by transformations like Stone-Wales transformation. (a) Two pentagons and central hexagon before $C_2$ insertion, (b) $C_2$ insertion leading to addition of new hexagon, (c) Pyracylene / Pyracene patch which can undergo SW transformation, (d) Pyracylene / Pyracene patch after SW transformation.

Figure 2. Growth of $C_2$ into $C_{60}$ by $C_2$ insertion, for n=2 to 20 energies of $C_{20}$ fullerene, graphene and corannulene growth is presented and for n > 20 $C_{60}$ energies of growth by open-caged fragments and close-caged fullerenes route is presented. All energies are calculated using MNDO method.

Figure 3. Growth of $C_{20}$ Corannulene, Fullerene and Graphene is modeled from $C_4$ up to $C_{20}$. HoF per carbon atom as a function of number of carbon atoms is also presented.

Figure 4. Growth of $C_{60}$ from $C_{24}$ is modeled by both open-caged fragments and close-caged fullerene routes. HoF per carbon atom as a function of number of carbon atoms is also presented.

Figure 5. Predicted growth path of $C_{24}$ fragment to $C_{24}$ Fullerene Transformation. HoF per carbon atom is presented for all 13 molecules. Transformation type during each molecular transformation is also presented.

Table 1. Parameters used in Semi-empirical Methods.

Table 2. Benchmarking of MNDO, AM1 and PM3 for $C_{60}$ Fullerene.

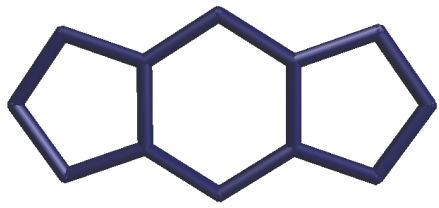 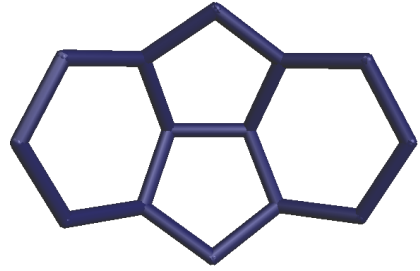

(a)  (b)

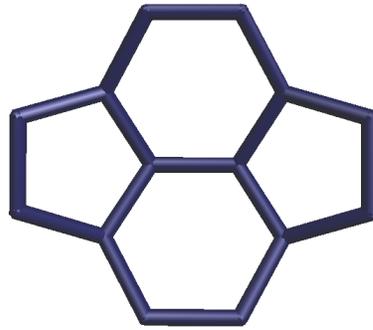 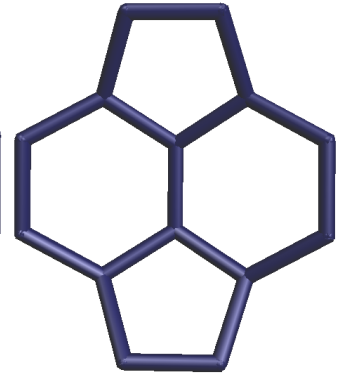

(c)  (d)

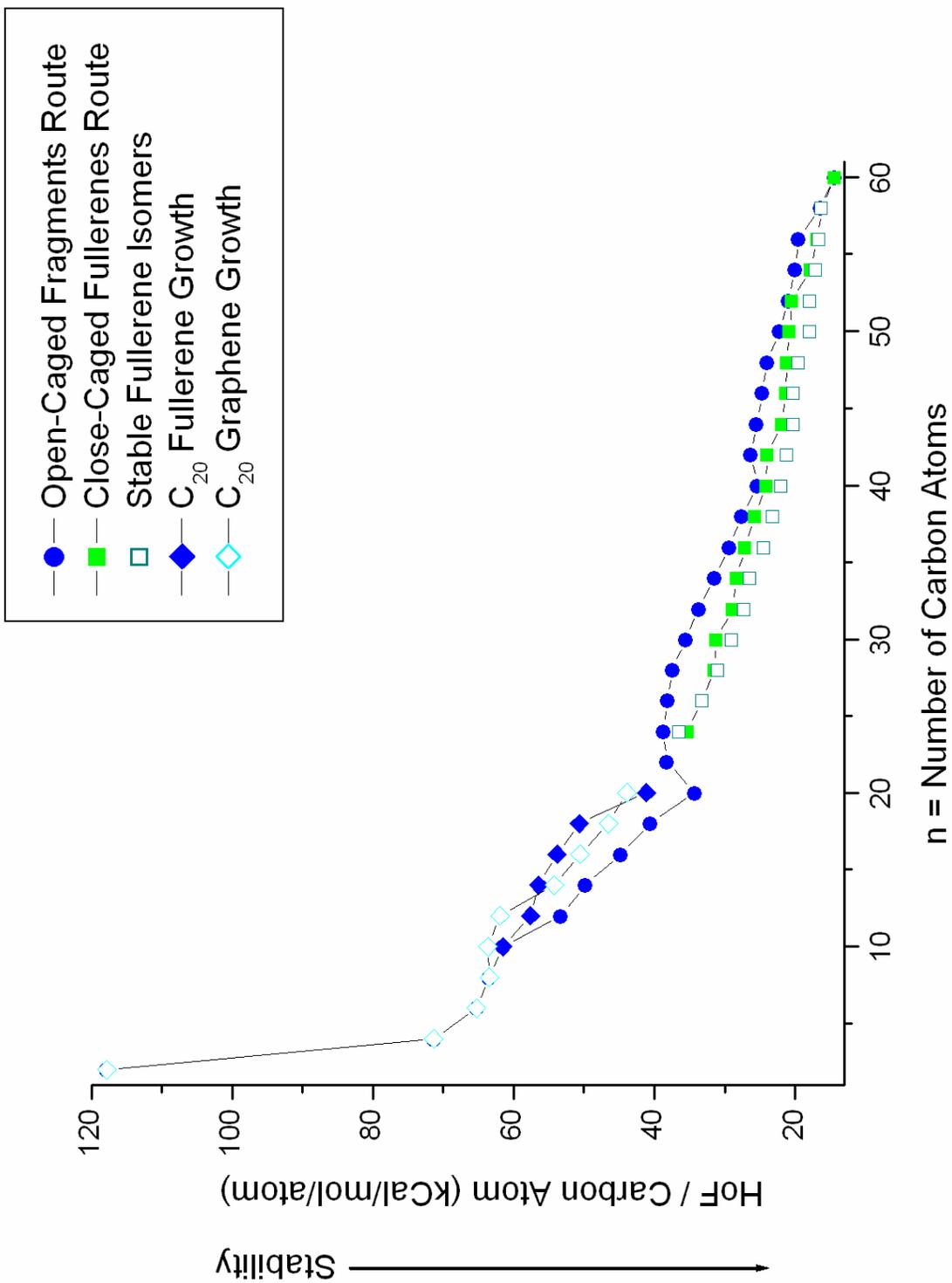

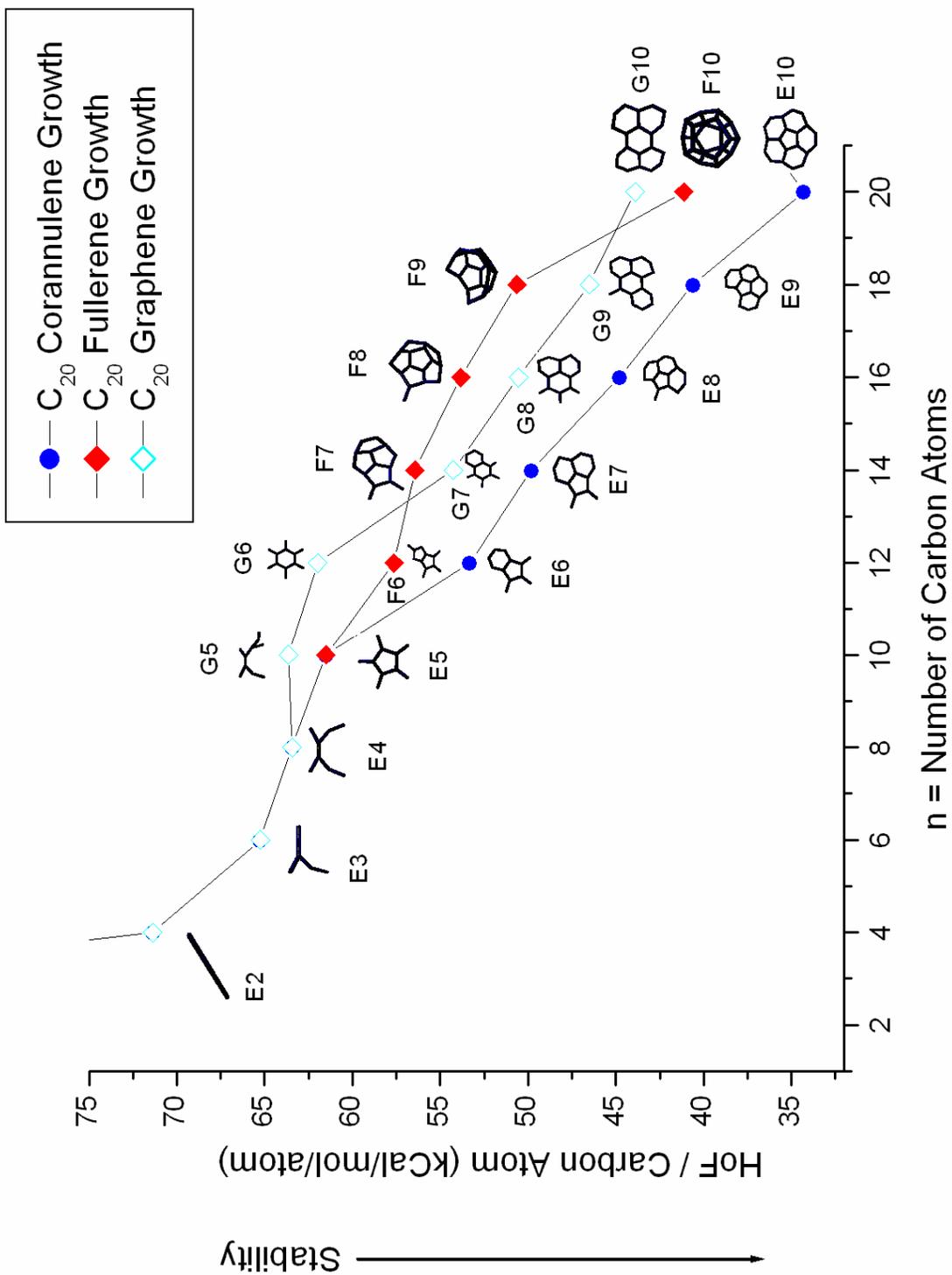

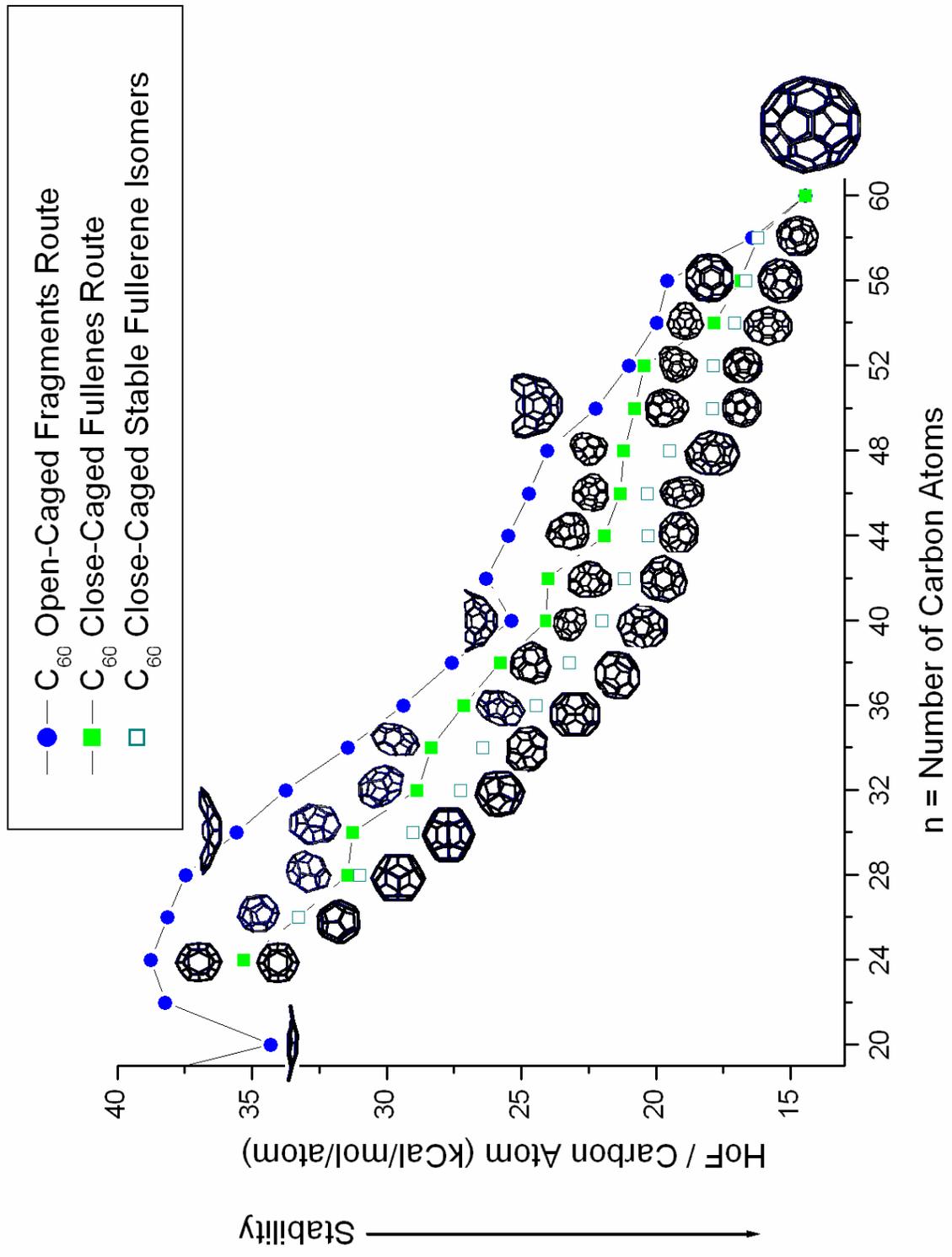

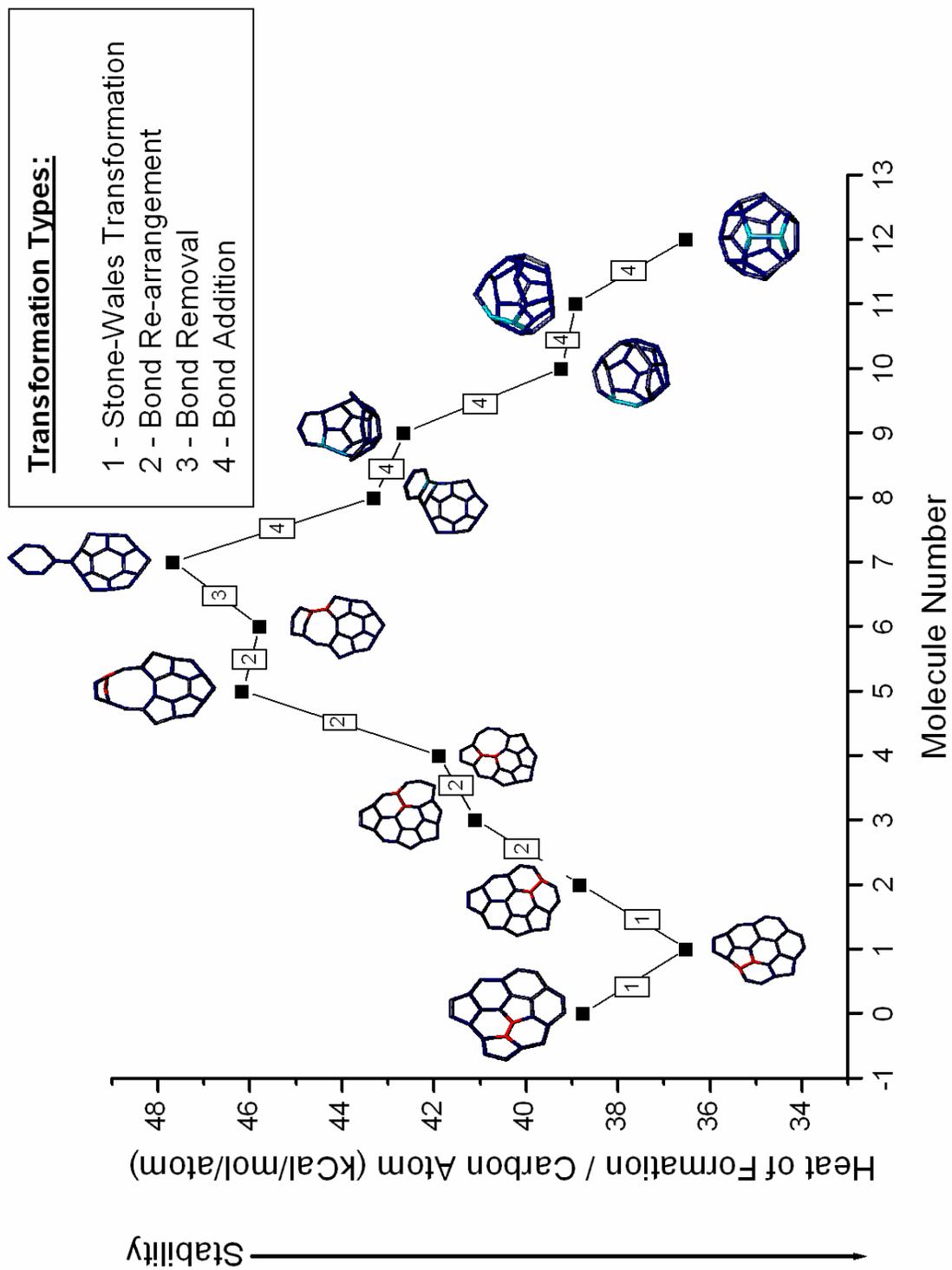
16

**Table 1**

| Parameter | Description | MNDO | AM1 | PM3 |
|---|---|---|---|---|
| $U_{ss}$ and $U_{pp}$ | S and p atomic orbitals; one-electron one-center integral | Optimized | Optimized | Optimized |
| $\beta_s$ and $\beta_p$ | s and p atomic orbital one-electron two-center resonance integral terms | Optimized | Optimized | Optimized |
| $\xi_s$ and $\xi_p$ | s and p-type Slater atomic orbital exponent | Optimized | Optimized | Optimized |
| $\alpha_A$ | Atom A core-core repulsion term | Optimized | Optimized | Optimized |
| $G_{ss}$ | s-s atomic orbital one center two electron repulsion integral | Experimental | Experimental | Optimized |
| $G_{sp}$ | s-p atomic orbital one center two electron repulsion integral | Experimental | Experimental | Optimized |
| $G_{pp}$ | p-p atomic orbital one center two electron repulsion integral | Experimental | Experimental | Optimized |
| $G_{pp'}$ | p-p' atomic orbital one center two electron repulsion integral | Experimental | Experimental | Optimized |
| $H_{sp}$ | s-p atomic orbital one center two electron exchange integral | Experimental | Experimental | Optimized |
| $K_{nA}$ or $a_{nA}$ | A Gaussian multiplier for $n^{th}$ Gaussian of atom A | - | Optimized | Optimized |
| $L_{nA}$ or $b_{nA}$ | A Gaussian exponent multiplier for $n^{th}$ Gaussian of atom A | - | Optimized | Optimized |
| $M_{nA}$ or $c_{nA}$ | A radius of center of $n^{th}$ Gaussian of atom A | - | Optimized | Optimized |

**Table 2**

| Method | Single Bond (Å) | Error (%) | Double Bond (Å) | Error (%) | Norm |
|---|---|---|---|---|---|
| Experimental [9] | 1.46 | | 1.4 | | |
| MNDO | 1.475 | 1.027397 | 1.4 | 0 | 1.055545 |
| AM1 | 1.465 | 0.342466 | 1.385 | -1.07143 | 1.265242 |
| PM3 | 1.459 | -0.06849 | 1.384 | -1.14286 | 1.310814 |